# Tunable interfacial chemisorption with atomic-level precision in a graphene/WSe$_2$ heterostructure

Mo-Han Zhang[1,†], Fei Gao[2,3,†], Aleksander Bach Lorentzen[2], Ya-Ning Ren[1], Ruo-Han Zhang[1], Xiao-Feng Zhou[1], Rui Dong[1], Shi-Wu Gao[4], Mads Brandbyge[2], Lin He[1,*]

[1]Center for Advanced Quantum Studies, Department of Physics, Beijing Normal University, Beijing, 100875, China, and Key Laboratory of Multiscale Spin Physics, Ministry of Education, Beijing, 100875, China

[2]Department of physics, Technical University of Denmark, 2800 Kongens Lyngby, Denmark

[3]Donostia International Physics Center (DIPC), 20018 Donostia-San Sebastián, Spain

[4]Beijing Computational Science Research Center, 100193 Beijing, China

[†]These authors contributed equally to this work.

[*]Correspondence and requests for materials should be addressed to Lin He (e-mail: helin@bnu.edu.cn).

**It has long been an ultimate goal to introduce chemical doping at the atomic level to precisely tune properties of materials. Two-dimensional materials have natural advantage because of its highly-exposed surface atoms, however, it is still a grand challenge to achieve this goal experimentally. Here, we demonstrate the ability to introduce chemical doping in graphene with atomic-level precision by controlling**

**chemical adsorption of individual Se atoms, which are extracted from the underneath WSe$_2$, at the interface of graphene/WSe$_2$ heterostructures. Our scanning tunneling microscopy (STM) measurements, combined with first-principles calculations, reveal that individual Se atoms can chemisorbed on three possible positions in graphene, which generate distinct pseudospin-mediated atomic-scale vortices in graphene. We demonstrate that the chemisorbed positions of individual Se atoms can be manipulated by STM tip, which enables us to achieve atomic-scale controlling quantum interference of the pseudospin-mediated vortices in graphene. This result offers the promise of controlling properties of materials through chemical doping with atomic-level precision.**

Chemical doping has been demonstrated to be one of the most effective ways to tune properties of materials. Manipulating the chemical doping with atomic-level precision would play a significant role in realizing the ultimate functional materials for electronics and optoelectronics devices. Two-dimensional (2D) materials facilitate a natural advantage for chemical doping at atomic level precision because all the atoms are exposed on the surface[1,2]. In the past few years, different strategies have been developed to introduce chemisorbed atoms or atomic defects in graphene to tune its properties[3-19]. It has been demonstrated explicitly that even a single chemisorbed H atom can drastically modify the magnetic and electronic properties of graphene. An outstanding example is that a single H atom adsorbed on graphene can result in a pseudospin-mediated atomic-scale vortex with angular momenta reflecting the Berry phase of graphene[16-18]. Although many great efforts have been made, precise control of chemisorption sites of individual atoms has not yet been achieved in experiment, owing to the limited precision of controlling the chemisorbed atoms.

In this work, we report a facile method to realize atomic-level precise doping in graphene through manipulating chemisorption of individual selenium (Se) atoms at the interface of graphene/$WSe_2$ heterostructures. Our scanning tunneling microscopy (STM) measurements, supported by first principles calculations, show that a single Se atom separated from $WSe_2$ can form bonds with carbon atoms in graphene at three different sites, A sublattice, B sublattice and the carbon-carbon (C-C) bridge, which introduce distinct pseudospin-mediated atomic-scale vortices in graphene. By using STM tip, we demonstrate the ability to manipulate the chemisorption positions of individual Se atoms on graphene and further obtain their possible combinations with atomic-level precision. This enables us to explore quantum interference of the pseudospin-mediated atomic-scale vortices in graphene. Our work opens a new avenue to tune properties of 2D materials through atomic-level precise chemical doping.

We chose the graphene/$WSe_2$ heterostructure because the individual Se atoms can be extracted from the $WSe_2$ while retaining the structural integrity of graphene on the top layer, and the interfacial Se atoms further chemisorbed on graphene by forming the Se-C bonds[20-22]. STM tip pulses were used to break the ionic bonds between Se and tungsten in the $WSe_2$. Our experiments indicate that conventional STM has sufficient spatial resolution to tune and characterize the interfacial Se-C bonds. Figure 1a schematically illustrates the process that an individual Se atom is extracted from the

WSe$_2$ substrate by applying a small voltage pulse and, simultaneously, a Se vacancy is generated at the position where the tip pulse is applied. Figure 1b shows a representative result obtained in the experiment and the interfacial chemisorbed Se atom, shown as a bright protrusion with apparent height ~157 pm, is about 2 nm away from the Se vacancy in the WSe$_2$ substrate, seen as an atomic-scale pit with depth about -67 pm. Our scanning tunneling spectroscopy (STS), *i.e.*, d$I$/d$V$, measurements of the Se vacancy show a localized electronic state at about 0.4 eV (see Fig. S1 of the Supplemental Material), as observed in the Se vacancy of the WSe$_2$ in previous studies[23,24]. According to our experiment, there is no broken bond in graphene generated by the tip pulses because that the C-C bond in graphene is much stronger than the Se-W bond in the WSe$_2$[25]. However, the atomic-scale large potential caused by the chemisorbed Se and Se vacancy can generate strong intervalley scattering in graphene, characterized as a threefold √3×√3 pattern that is rotated 30° degrees with respect to the graphene lattice, as shown in Fig. 1b.

The chemisorbed Se atoms at the interface exhibit two kinds of distinct features in the STM images, as shown in Figs. 1d and 1h respectively. The most-frequently observed case is that a pair of the nearest C atoms become the highest sites in the STM image and the induced intervalley scattering exhibits a mirror symmetry about a C-C bond (Fig. 1d). Such characteristic features are similar to that of a nitrogen atom chemisorbed on a C-C bond in graphene[26] and the observed pattern is attributed to the chemisorption of an individual Se atom at the bridge site in graphene. Since there are three directions of the C-C bonds in graphene, we can observe three directions of such pattern. The other case, as depicted in Fig. 1h, has a three-fold symmetry with the highest site localized at the center of a C atom in graphene. This is similar to that of a single atom absorbed on C atom in graphene[7,9,10] and the observed pattern is associated with the chemisorption of an individual Se atom on graphene sublattice. To gain a deep insight, we carry out a series of multi-scale atomistic simulations on a single Se atom chemisorbed on graphene at the bridge and C atom sites. For simplicity, we don't consider the influence of the $WSe_2$ substrate, which is not expected to affect the main features observed here due to the weak vdW interaction between graphene and the $WSe_2$. The local atomic structures in both two cases exhibit the out-of-plane buckling of the C atoms around the chemisorbed Se atom, as shown in Figs. 2a and 2e. Obviously, the simulated local density of states (LDOS) for the two adsorption sites are in good agreement with that measured in experiments. Our theoretical calculations also reveal that the most stable state is the Se at the bridge site with the binding energy of ~1.02 eV, and it costs a modest energy expense of 0.17 eV when the Se is at the C site. This suggest the structure that the Se atom chemisorbed on the C-C bridge is easier to obtain, which is in line with the experiments.

The chemisorbed Se atom is expected to generate localized electronic states near the Fermi level ($E_F$) in graphene. Figures 1g and 1k show the spatially resolved STS contour plots along the yellow arrows in Figs. 1b and 1h, respectively. Apparently, the d$I$/d$V$ spectra exhibit a peak at 280 meV when the Se atom is chemisorbed at the C-C bridge and display a peak at 180 meV when the Se atom is chemisorbed on the C atom. The localized states extend several nanometers away from the chemisorbed Se atom and the weaker peaks in the spectra recorded away from the chemisorbed Se atom are attributed to tip-induced quasibound states in pristine graphene monolayer, as observed previously[27,28]. To understand the origin of the localized states, we analyze the low-energy projected DOS (PDOS) for the two adsorption sites, as shown in Figs. 2d and 2k, respectively. A peak originated from the hybridization between Se-$p_z$ and C-$p_z$ orbitals is observed at 500 meV above the $E_F$ for the ground state, and a similar peak appears at 200 meV when the Se atom is chemisorbed on the C site, which is qualitatively in line with our experimental results. The concentration of the chemisorbed Se atoms in the simulations is greater than that in the experiments, which may cause the higher energy of the resonance peaks seen in the theoretical calculations. The neglect of the substrate and tip in theory may also partially contribute to the difference between the experimental and theoretical results. Thus, we demonstrate explicitly that the individual Se atoms extracted from the WSe$_2$ can chemisorbed on graphene at the interface of the graphene/WSe$_2$ heterostructures.

Although all the chemisorbed Se atoms can generate strong intervalley scattering in graphene, as shown in both the STM images (Figs. 1d and 1h) and their Fourier transform (FT) images (inner bright spots connected by the red dashed hexagon in the FT images of Figs. 1e and 1i), the induced intervalley scatterings exhibit distinct pseudospin-mediated physics. Figures 1f and 1j show inverse FT-filtered images for two different Se adsorption sites, which exhibit different characteristics. No dislocation ($|N| = 0$) is observed in the vicinity of the Se atom adsorbed at the C-C bridge in graphene, whereas two additional wavefronts dislocation (marked by yellow lines in

Fig. j, $|N| = 2$) appear when the Se atom is chemisorbed on the C site. The chemisorbed Se atom will introduce both intravalley scattering and intervalley scattering in graphene, as schematically shown in Fig. 1c. The intravalley scattering involves a rotation of the momentum-locked pseudospin that is always π, as shown in the right panel of Fig. 1c. Thus, the interference is counterbalanced at the leading order. In contrast, the intervalley backscattering involves a rotation of the pseudospin by an angle $-2\theta_q = -2\theta_r$ (see the left panel in Fig. 1c)[16-18]. An accumulation of the phase shift over a closed path enclosing the single atomic defect is $\pm \int_0^{2\pi} 2 d\theta_q = \pm 4\pi$. Motivated by this, a single defect at the A (B) sublattice of graphene can be regarded as a pseudospin-mediated atomic-scale vortex with angular momenta ±2. In our cases, the Se atom adsorbed at graphene A (B) sublattice has the angular momenta $l = +2$ ($l = -2$), reflecting the Berry phase in the monolayer graphene[16-18]. Meanwhile, the angular momentum of the Se atom chemisorbed at the C-C bridge is $l = +2 + (-2) = 0$, because the Se atom establish bonds with two C atoms which belong to the A and B sublattice in graphene, respectively. The FT-filtered STM images with the specific direction of intervalley scattering implies the interference patterns of a uniform plane wave and a defect induced atomic-scale vortex. Therefore, the total number of additional wavefronts reflects the angular momentum of the vortex in the interference patterns. The above analysis is further confirmed by the first principles calculations. The FT from the simulated LDOS in Figs. 2a and 2e also shows signal of the intervalley scattering (see Figs. 2b and 2f). Furthermore, the FT-filtered images, as shown in Figs. 2c and 2g, have the same number of the corresponding additional wavefronts dislocation as it in the experiments, surrounding the Se atom at the bridge ($|N| = 0$) and the C lattice ($|N| = 2$) in graphene, respectively (see Method part and SI for computational details). Both our experiments and theoretical calculations indicate that a Se atom chemisorbed at two different sites in graphene can introduce dramatically different electronic properties.

Besides in-situ creating of chemisorbed individual Se atoms on graphene, our experiment further achieves that the precise manipulation of the chemisorption positions of a single Se atom and chemical bonds between the Se and graphene using the electrical field in STM. Figure 3 summarizes a representative example. With the help of a reference object, as outlined with a white dotted triangle in Figs. 3a, 3c, and 3e, our experiment demonstrates explicitly that we can tune the chemisorbed positions and bond between the Se atom and graphene. In Figs. 3a and 3e, the Se atom chemisorbed on the A sublattice and B sublattice, respectively, which can be identified thanks to the different orientations of the Se-induced tripod shapes in the STM images. In Fig. 3c, the Se atom chemisorbed on the C-C bridge, exhibiting the same features shown in Fig. 1d and Fig. 2a. The tunable chemisorbed positions of the Se atom change the pseudospin-mediated intervalley scattering around them. Figures 3b, 3d, and 3f show the corresponding FT-filtered images, where the $N = 2, 0, -2$ additional wavefront dislocations are obtained in the vicinity of the chemisorbed Se atom, respectively (see Fig. S2 for the original STM images), indicating the variation of angular momenta $l = +2$, $l = 0$, $l = -2$ of the pseudospin-mediated vortices. Thus, the electronic properties of graphene can be effectively modulated by controlling adsorption sites with atomic-level precision in the experiments. In addition, according to previous studies[7-9,11-14], a spin splitting of the C-$p_z$ peaks around the $E_F$ causes a local magnetic moment in graphene, and such spin moment can be manipulated by the local field via a STM tip[29]. Therefore, we can possibly tune both pseudospin-mediated and spin-related properties in graphene by manipulating the chemisorption sites of individual Se atoms (See Fig. S3 for more experimental results of manipulation the interfacial Se atoms) and the filling of the localized states of adsorbed Se atoms.

With the ability to control the chemical reaction between the individual Se atom and graphene with atomic-level precision, then, it becomes possible to combine different kinds of chemisorbed Se atoms to further tune properties of graphene. Figure 4 summarizes two representative results that two individual Se atoms are chemisorbed at opposite (AB chemisorption) or the same sublattices (AA chemisorption) in graphene (See Fig. S4 for more experimental results). The chemisorbed positions are identified according to the different orientations of the tripod shapes in the STM images. For the case of the AB chemisorption (Fig. 4a), there is zero additional wavefront dislocation, as shown in Fig. 4b, resulting from the annihilation of dislocations of the pseudospin-mediated vortices with opposite angular momenta. For the case of the AA chemisorption (Fig. 4c), it is interesting that the number of additional wavefronts induced by the two A-site adsorptions is still 2 (Fig. 4d). Such a counterintuitive result can be well explained with considering the interference of two pseudospin-mediated vortices[17]. The winding number of the pseudospin over a closed path surrounding two $l = +2$ vortices is still 2, consequently, we obtain $|N| = 2$ additional wavefronts. With considering the fact that individual atoms chemisorbed on A and B sublattices of graphene will result in local magnetic moments in opposite directions[7,30-32], it is expected to obtain completely different magnetic coupling between the local magnetic moments for the AB and AA chemisorption. In the near future, by controlling large-area chemisorption of the Se atoms only on the A (or B) sublattice, it has highly potential to realize 100% spin polarization in graphene.

In summary, we demonstrate the ability of extraction of individual Se atoms and manipulation of its chemisorption on graphene with atomic-level precision. By manipulating the adsorption sites of individual Se atoms, our experiments supported by our theoretical calculation, indicate that we can introduce distinct pseudospin-mediated atomic-scale vortices in graphene and explore their quantum interference. The reported *ab-initio* approach combines high spatial resolution of STM technique and the highly tunable chemical bonding between individual Se atoms and graphene. Our study lays a solid foundation for realizing custom-designed future materials through atomic-level precise chemical doping.

**Data availability**

The data that support the plots within this paper and other findings of this study are available from the corresponding authors upon reasonable request.

**Acknowledgments:**

This work was supported by the National Key R and D Program of China (Grant Nos. 2021YFA1401900, 2021YFA1400100), National Natural Science Foundation of China (Grant Nos.12141401, 11974050, 11934003, U1930402), and "the Fundamental Research Funds for the Central Universities" (Grant No. 310400209521). F.G. and M.B. acknowledge support from Villum fonden (Project No. VIL13340), and A.B.L. acknowledges support from the Independent Research Fund Denmark (Project No. 0135-00372A).

**Author contributions**

M.H.Z. performed the sample synthesis, characterization and STM/STS measurements. F.G. and A.B.L. performed the DFT calculations. L.H. conceived and provided advice on the experiment, analysis and the theoretical calculations. M.H.Z., Y.N.R., R.H.Z., X.F.Z., R.D. and L.H. acquired and analyzed the experimental data. F.G., A.B.L., S.W.G., and M.B. analyzed the theoretical calculations. M.H.Z., F.G. and L.H. wrote the manuscript. All authors discussed the results and commented on the manuscript.

# Figures

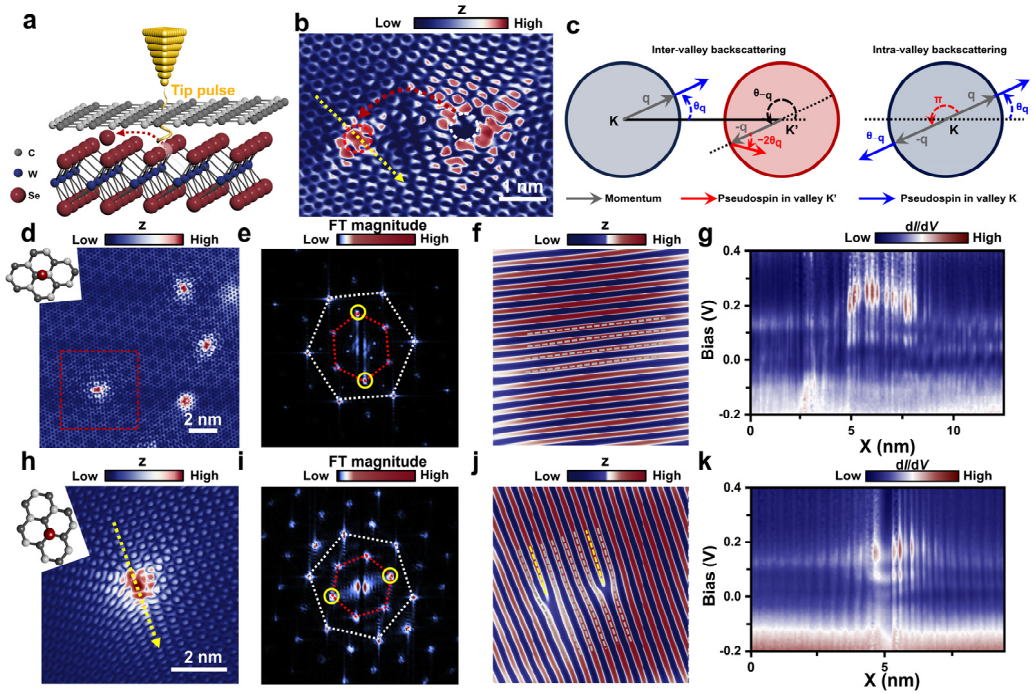

**Fig. 1 | Interfacial chemical reactions between an individual Se atom and graphene.**

**a**, Schematics of a graphene/WSe$_2$ heterostructure after a tip pulse. The STM tip pulse creates a Se adatom and a Se vacancy in WSe$_2$. **b**, STM topography of a graphene/WSe$_2$ heterostructure with a single Se atom chemisorbed on monolayer graphene and a Se vacancy in the WSe$_2$ substrate ($V_b$= -0.2 V, $I$= 200 pA). The Se adatom and Se vacancy are marked by red and white dotted circles respectively. The red dotted line is a possible trajectory of the Se atom. **c**, The backscattering process in graphene. Inter-valley backscattering between wavevector states **q** and -**q** belonging to the valleys K and K′, which leads to a rotation of -$2\theta_\mathbf{q}$ of the pseudospin. Intra-valley backscattering rotates the pseudospin by π. **d** and **h**, Topography STM images of a Se adatom under the C-C bridge and the carbon atom of the graphene respectively. In panel d, $V_b$ = 0.07 V and $I$ = 200 pA. In panel h, $V_b$ = 0.1 V and $I$ = 200 pA. The insets present atomic structures for the Se atoms (red balls) chemisorbed on graphene (light and dark gray balls). For clarity, the Se atom is plotted on top of graphene. **e**, Modulus of the FT of the image in red square of **d**. **i**, Modulus of the FT of the image in **h**. In panels **e** and **h**, the outer hexangular spots (corners of the white dotted line) and inner bright spots (corners of the red dotted line) correspond to the reciprocal lattice of graphene and the interference of the intervalley scattering, respectively. **f, j**, FT-filtered images enclosed by yellow circles in (**d, h**) along one of the three directions of inter-valley scattering. The yellow dashed lines correspond to $|N| = 2$ additional wavefronts. **g, k,** Spatially resolved contour plots of d$I$/d$V$ spectra along the yellow arrows in (**b, h**), respectively. The locations of the Se atom are at $x = 7$ nm and 5 nm, respectively.

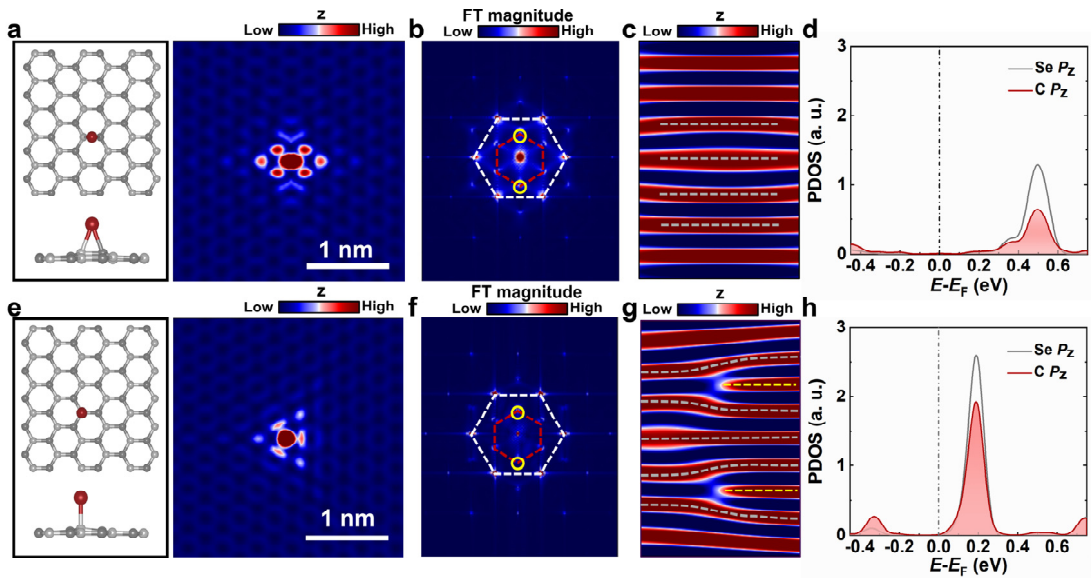

**Fig. 2 | Simulated electronic properties of a single Se atom chemisorbed on graphene. a** and **e,** Optimized atomic geometries (left panels) and LDOS (right panels) of a Se adatom at the C-C bridge and the C atom of the surface of graphene, respectively. Red, light gray and dark gray balls represent Se atom, C atoms in A and B sublattice, respectively. **b, f,** The FT images of the LDOS. The outer hexangular spots (corners of the white dotted lines) and inner bright spots (corners of the red dotted lines) correspond to the reciprocal lattice of graphene and the interference of the intervalley scattering, respectively. **c, g,** FT-filtered images enclosed by yellow circles in (**b, f**) along one of the three directions of inter-valley scattering. The yellow dashed lines correspond to $|N| = 2$ additional wavefronts. **d, k,** The projected DOS on the Se atom and C atoms near the chemisorbed Se. Black and red lines show Se-$p_z$ and C-$p_z$ orbitals, respectively.

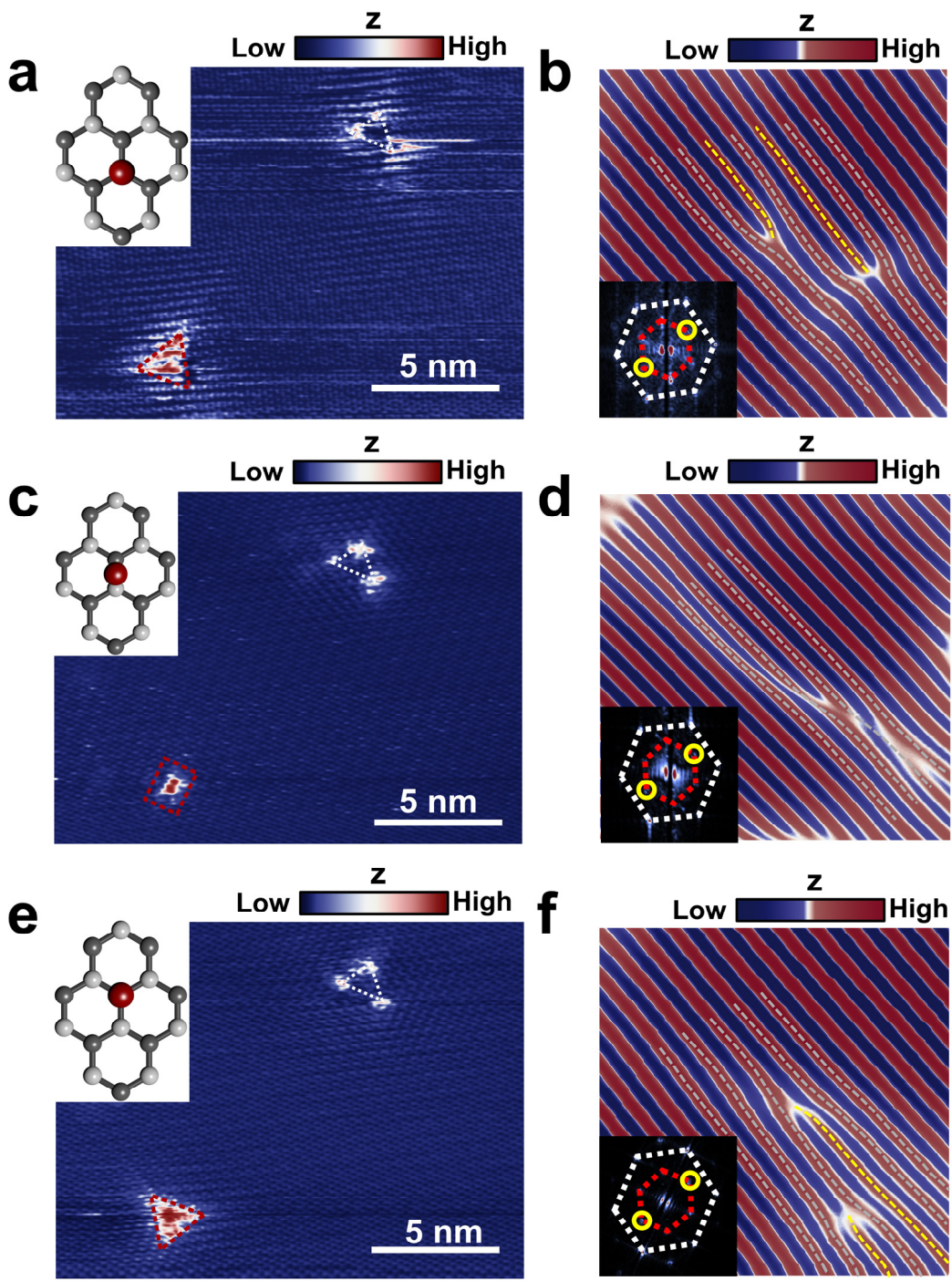

**Fig. 3 | Manipulating the chemisorption of individual Se atom on graphene. a**, **c**, **e**, STM images of a single Se adatom chemisorbed on A sublattice, C-C bridge and B sublattice of graphene ($V_b = 0.4$ V, $I = 200$ pA), respectively. The defect, as marked with dotted white lines, is used as a reference object. The schematic atomic structures are given in the insets. For clarity, the Se atom is plotted on top of graphene. The red

triangle indicates the different orientations of the tripod shapes induced by the chemisorbed Se atom. **b, d, f**, FT-filtered images near the Se adatom shown in **a, c, e** along the marked pair of signatures of intervalley scatterings. Insets: FT of the STM images (see Fig. S2 for the original STM images). Graphene and intervalley scattering Bragg peaks are connected with white and red dotted lines, respectively. The yellow dashed lines correspond to $N = +2, 0, -2$ additional wavefronts, respectively.

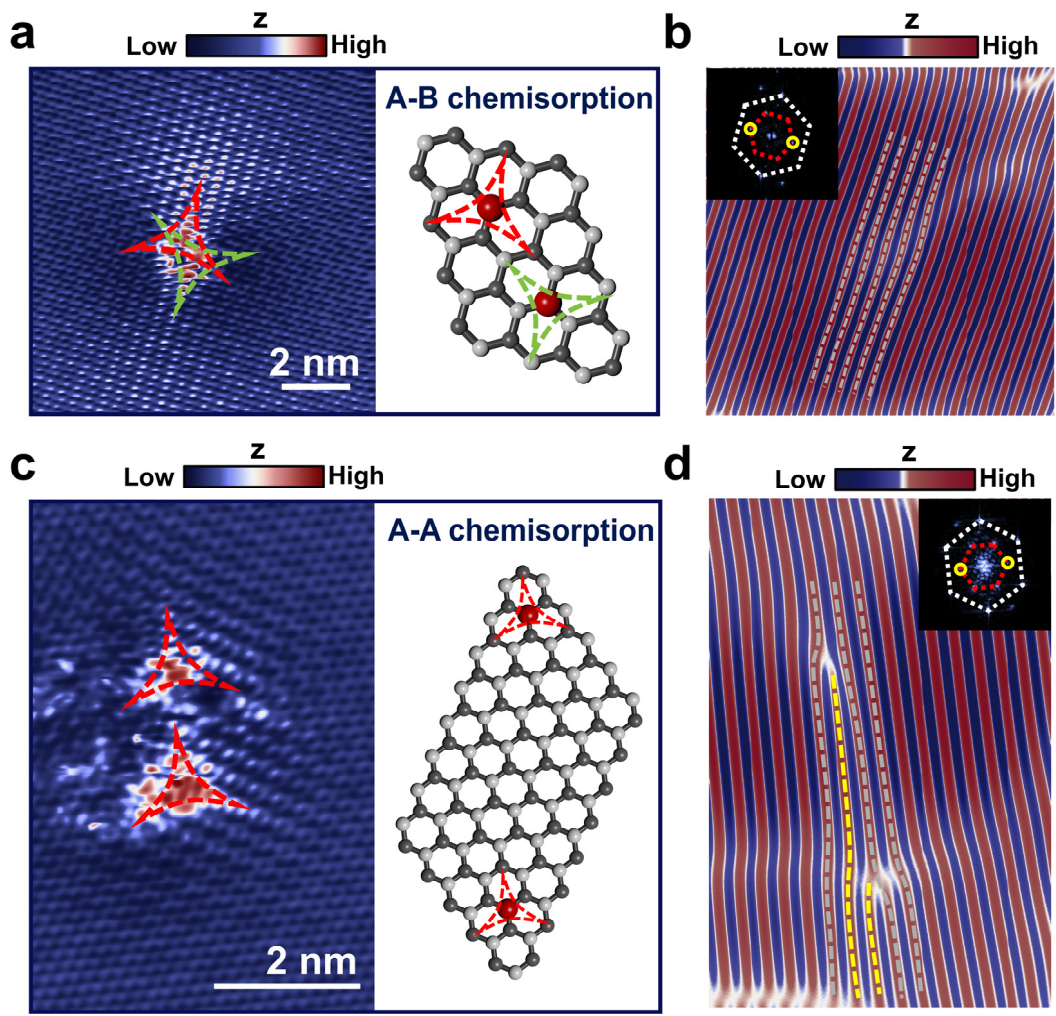

**Fig. 4 | Interference of pseudospin-mediated vortices and wavefront dislocations induced by A-B chemisorption and A-A chemisorption. a**, **c**, Typical STM images of two individual Se adatoms chemisorbed on different or the same sublattices of graphene, respectively ($V_b$ = 0.15 V, $I$ = 200 pA in **a**, $V_b$ = 0.18 V, $I$ = 200 pA in **b**). The dotted tripod shapes are added manually to indicate the orientations and positions of the defects. The right panels present the corresponding atomic structures of the Se atoms (red balls) absorbed on graphene. **b**, **d**, FT-filtered images in **a**, **c** along the marked reciprocal lattice points of intervalley scatterings. Insets: the filters applied in the Fourier space. White and red dotted lines connect the graphene and intervalley

scattering Bragg peaks, respectively. The yellow dashed lines correspond to $|N| = 2$ additional wavefronts in **d**.

**Method**

**Samples preparation.** The studied devices were fabricated with a wet-/dry-transfer technique using the transfer platform from Shanghai Onway Technology Co., Ltd. First of all, the WSe$_2$ crystal was separated into thick-layer WSe$_2$ sheets by scotch tape. We stacked thick-layer WSe$_2$ sheets to 285 nm thick SiO$_2$/Si with polydimethylsiloxane (PDMS). The Si in the bottom is highly N-doped. Then, we grow large area graphene monolayer films on a 20 × 20 mm$^2$ polycrystalline copper (Cu) foil (Alfa Aesar, 99.8% purity, 25 μm thick) via a low-pressure chemical vapor deposition (LPCVD) method. The Cu foil was heated from room temperature to 1035 ºC in 30 min and annealed at 1035 ºC for ten hours with Ar flow of 50 sccm and H$_2$ flow of 50 sccm. Then CH$_4$ flow of 5 sccm and the other gases to be the same as before was introduced for 20 min to grow high-quality large area graphene monolayer. The furnace was cooled down naturally to room temperature. We used wet-transfer technique with polymethyl methacrylate (PMMA) to transfer graphene monolayer onto the substrate (WSe$_2$/SiO$_2$/Si). Finally, we cleaned the upper PMMA with acetone.

**Measurements**. STM/STS measurements were performed in low-temperature (78 K) and ultrahigh-vacuum (~10$^{-10}$ Torr) scanning probe microscopes [USM-1300 (78 K)] from UNISOKU. The tips were obtained by chemical etching from a W (99.95%) alloy wire. Before the experiment, we calibrate the STM with Highly Oriented Pyrolytic Graphite. The differential conductance (d$I$/d$V$) measurements were taken by a standard lock-in technique with an ac bias modulation of 5 mV and 793 Hz signal added to the tunneling bias.

**Density functional theory**. The first principles calculations were performed using the SIESTA code with the GGA-PBE[33] functional for exchange-correlation. We adopted a standard double-ζ basis with polarization orbitals (DZP)[34-36]. The energy cutoff of 400 Ry was set to define the real-space grid for numerical integration of electron density. The K-point mesh of 9 × 9 × 1 was used for a 9 × 9 graphene cell and all atoms were allowed to relax until the forces on each atom have magnitudes less than 0.02 eV/Å. In addition, a 20 Å thick vacuum layer for the slab model was introduced. The electronic structure calculations were checked by comparing to plane-wave calculations[37](See Fig. S5 of the Supplemental Material). The FT and FT-filtered images were simulated by extending the system cell to a 45 × 45 graphene cell with open boundary condition[38] in a multi-scale calculation[39]. Physical quantities like local density of states were extracted, and accurate "pruned" DFT Hamiltonians for large-scale simulations were constructed using SISL[40](See Fig. S6 of the Supplemental Material).